\documentclass[twocolumn,showpacs,preprintnumbers,amsmath,amssymb,prb,aps]{revtex4-1}
\usepackage{epsfig}
\usepackage{epstopdf}
\usepackage{graphicx}
\usepackage{dcolumn}
\usepackage{bm}
\usepackage{textcomp}
\usepackage{array}
\usepackage{subcaption}
\usepackage{booktabs}
\usepackage{blkarray} 
\usepackage{amsmath}
\usepackage[mathscr]{euscript}
\usepackage[utf8x]{inputenc}

\begin{document}

\title{First-principles evidence of type-II Weyl phonons in rock-salt Tin Chalcogenides (SnS, SnSe \& SnTe) materials}

\author{Antik Sihi$^{1,}$}
\altaffiliation{sihiantik10@gmail.com}
\author{Sudhir K. Pandey$^{2,}$}
\altaffiliation{sudhir@iitmandi.ac.in}
\affiliation{$^{1}$School of Basic Sciences, Indian Institute of Technology Mandi, Kamand - 175075, India\\
$^{2}$School of Engineering, Indian Institute of Technology Mandi, Kamand - 175075, India}

\date{\today}

\begin{abstract}
 Recent studies on different topological materials in condensed matter physics have provided the evidence of topological nature for bosonic particle like phonons by performing various theoretical calculations and experimental observations. Here, the topological behaviours of phonons of SnS, SnSe and SnTe materials in rock-salt structure are investigated using $ab$-$initio$ methodology. For all these materials, the tilted linear band touching is observed along with the presence of band inversion in direction X-W. The topological point in phonon dispersion curve along X-W direction is found to be $\sim$2.83, $\sim$2.46 and $\sim$2.51 THz for SnS, SnSe and SnTe, respectively. The calculated numbers of Weyl points (WPs) is estimated to be 56, 24 and 54 for SnS, SnSe and SnTe, respectively. These WPs have shown conserved Chiral charges for all corresponding materials. The surface local density of states is computed for these compounds, where the surface arc is clearly seen. In case of SnS, the surface arc is found at 2.0 - 2.2 THz energy region. Also, the isofrequency surface states are investigated to observe the presence of Fermi arc for SnX (X = S, Se, Te). The present first-principles calculation predicts that SnS, SnSe and SnTe show type-II Weyl phononic behaviour along with evidence of topological phononic surface states.  
\end{abstract}

\maketitle


 \textit{Introduction :-} Nowadays, topological materials (TM) bring new era of research in the field of condensed matter physics because they have shown different unique properties for example topological insulator, topological semimetal, quantum hall-effect $etc.$\cite{hasan_rev, bansil_rev, yan_rev, viswa_rev} All these properties give opportunity to the researcher for developing the new theory of individual field and have provided the possibility of various applications. In present days, the most emergent applications of different topological properties are known as valleytronics \cite{valley}, spintronics based devices \cite{spin}, quantized circular photogalvanic effect \cite{photogal} $etc$. It is known that the bulk electronic band structure of any topological insulator usually show band inversion near the bandgap and the conducting surface states of this corresponding material are typically protected by time-reversal symmetry. But, in case of topological semimetals for fermionic system, the linear band touchings in three-dimensional (3D) crystal may host the excitation due to Weyl or Dirac fermions in 3D brillouin zone (BZ) \cite{viswa_rev,vivek}. In case of former one, the touching points of two non-degenerate bands are typically called as Weyl points (WPs). These points generally show non-zero topological charges based on the Nielsen-Ninomiya theorem \cite{nielsen}, which is known as the Chirality ($\mathscr{C}$ = +1 or -1) of the corresponding point. Each of these opposite chiral points represents the monopole of the Berry curvature. At a particular energy surface, the opposite two chiral points are typically connected through the topological surface state, which is known as the Fermi arc \cite{hasan_weyl}. Now, depending on the nature of linear band touching, the Weyl semimetals are sub-divided into type-I and type-II \cite{prl_117, nature_527}. In case of type-I, the usual conical band touching is observed, where point like Fermi surface is generally observed \cite{weng_5}. But, tilted Weyl cones are found for type-II Weyl semimetals. Further, it has been reported that the topological Lorentz invariance may be possible to be broken at the energy of the WPs, which may be responsible for these materials to show different exotic electromagentic responses \cite{tamai,Huang, chang, yu, udagawa}. TaAs is known as the first material in the list of type-I Weyl semimetal \cite{xu,lv}, whereas WTe$_2$ is crowned as the first realized material in the class of type-II Weyl semimetal \cite{nature_527, xu_115}. Moreover, the other examples of type-II Weyl fermionic materials are MoTe$_2$ \cite{Huang,wang}, WP$_2$, MoP$_2$\cite{autes}, TaIrTe$_4$ \cite{koepernik}, NbIrTe$_4$\cite{liu_17}, Ta$_3$S$_2$\cite{chang} $etc.$

  Till now, huge amount of theoretical and experimental works have been carried out to investigate the topological properties of fermionic electrons for many different materials. Phonon, which is known as the collective excitations of atoms within the sample, is a bosonic particle. It also shows many exotic phenomena like thermal conductivity \cite{guthrie, li}, mechanical properties\cite{tang}, phase transition\cite{fleury}, superconductivity \cite{bohnen,liu,zhang} etc. depending on the corresponding phonon modes of any compound. Recently, different non-trivial topological properties of these vibrational modes have been investigated both from theoretical and experimental observations \cite{miao, tzhang, ttzhang}. In case of this bosonic particle, the topological phenomena is generally sub-divided into Weyl phonons and Dirac phonons, depending on the symmetry conservations \cite{tzhang,chen}. Different chalcogenides based materials are already categorized as candidate of topological phonons \cite{tzhang,wang_95,li_97,xia_cdte,liu_103}. But, the list of type-II Weyl phononic sample is very small, which typically restricts the applicability of this type of material. Therefore, to find out more realistic type-II Weyl phononic materials, the first principle calculation is required to be carried out on different chalcogenides based compounds. In addition to this, if the electronic band structure of any material shows non-trivial topological nature along with the non-trivial phononic topological behaviour, then this particular material will obviously provide more ways for application purpose. Till our present knowledge, this aspect is not focused in any earlier work. To perform this work, tin chalcogenides (SnX, X = S, Se, Te) are the best candidate, which are possible to be experimentally synthesized and the host of fermionic topological behaviour in rock-salt phase \cite{antik_pla, asihi_sse, liu_snte, liu_snte, tanaka_snte, wang_snte,y_sun, wang_snse_exp1, jin_snse_exp2, mariano_exp_sns, bilenkii_exp_sns2, skelton_sns}. Moreover, the linear band touching is also observed from the phonon band structures of SnS and SnSe materials, which has motivated us to explore the phononic topological behaviour of these materials \cite{asihi_sse}.
  
  In this work, we have investigated the phononic topological behaviour of SnS, SnSe and SnTe materials in rock-salt structure. The linear band touching along X-W direction for all these mentioned materials are seen from respective phonon dispersion curves. The tilted nature of the band touching indicates the presence of type-II Weyl phonon at frequency value of $\sim$2.83 THz for SnS. The behaviour is more clearly seen both for SnSe and SnTe, where the touching points are situated at $\sim$2.46 THz and $\sim$2.51 THz for SnSe and SnTe, respectively. Moreover, the estimated values of WPs are found to be 56, 24 and 54 numbers for SnS, SnSe and SnTe, respectively, within the first full BZ. The values of Chirality are estimated to be +1 for half of WPs and -1 for the other half of WPs for all the materials. Moreover, the presence of surface arc is suggested from the surface local density of states. The isofrequency surface states are also computed for getting the information about the Fermi arc at particular frequency. The Fermi arc is seen at frequency $\sim$2.2THz and $\sim$2.25 THz for SnS material. The similar behaviour is also observed for SnSe and SnTe compounds.

\begin{figure}[]
    \begin{subfigure}{0.85\linewidth}
    \includegraphics[width=0.92\linewidth, height=4.0cm]{sns_ph.eps}
    \caption{}
   \label{fig:} 
\end{subfigure}
\begin{subfigure}{0.45\linewidth}
   \includegraphics[width=0.8\linewidth, height=2.5cm]{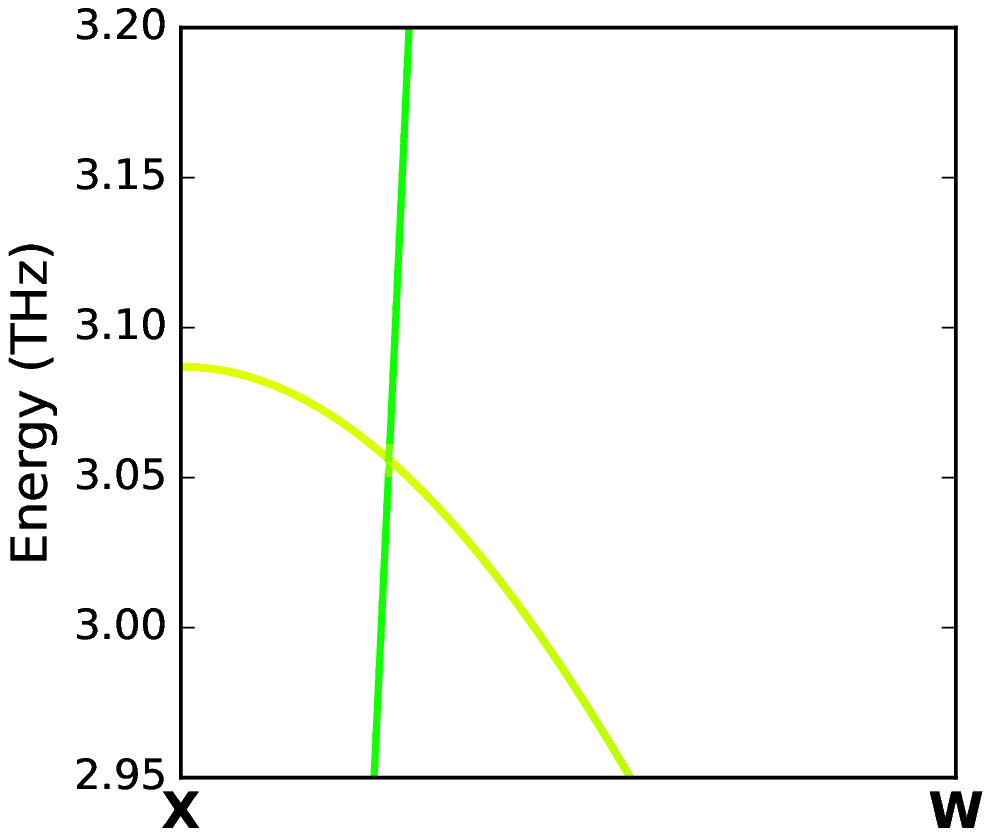}
   \caption{}
   \label{fig:}
\end{subfigure}
\begin{subfigure}{0.45\linewidth}
   \includegraphics[width=0.8\linewidth, height=2.5cm]{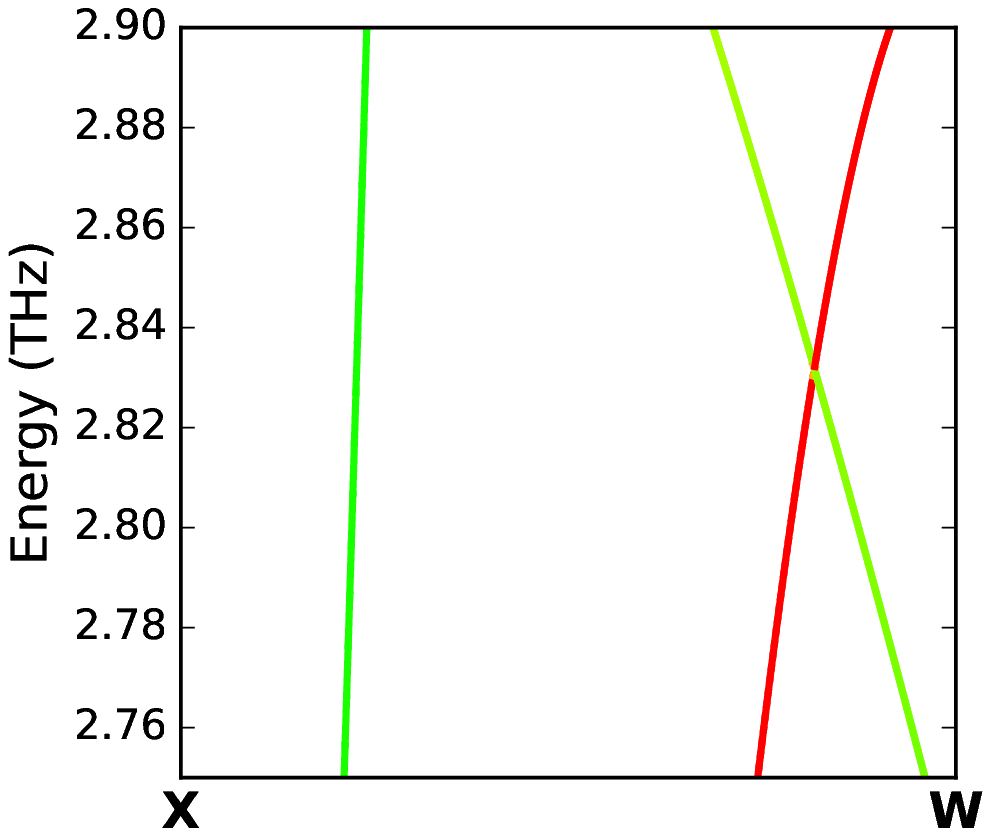}
   \caption{}
   \label{fig:}
\end{subfigure}
\begin{subfigure}{0.45\linewidth}
   \includegraphics[width=0.8\linewidth, height=2.5cm]{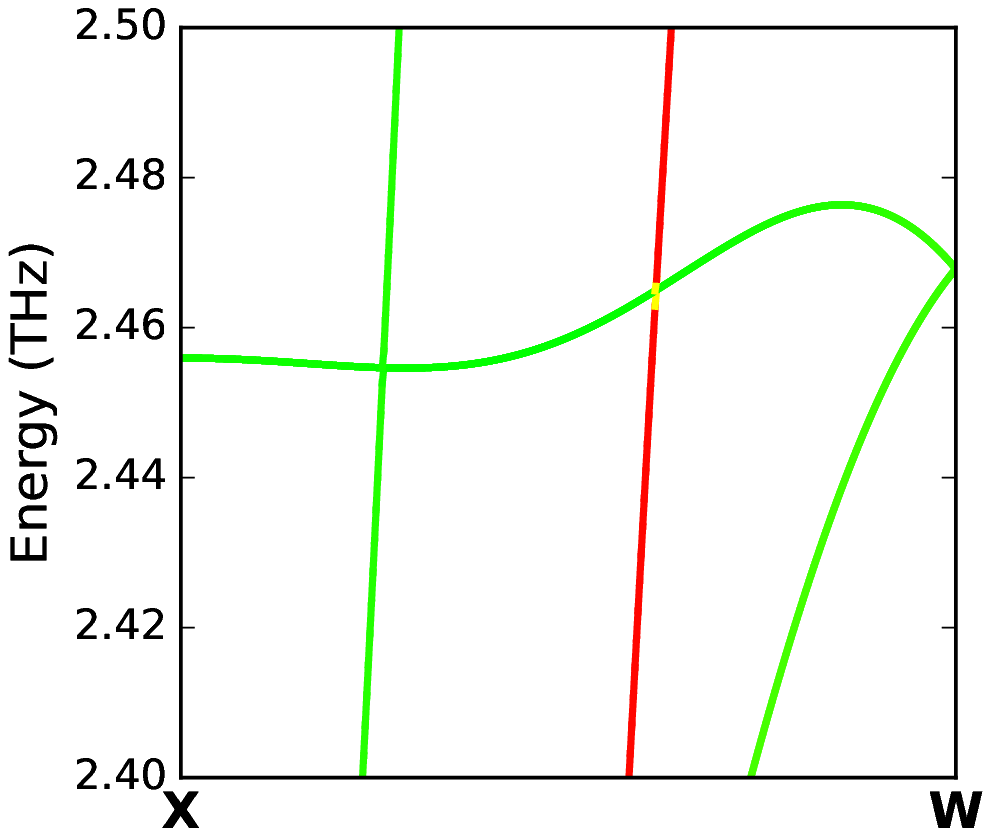}
   \caption{}
   \label{fig:}
\end{subfigure}
\begin{subfigure}{0.45\linewidth}
   \includegraphics[width=0.8\linewidth, height=2.5cm]{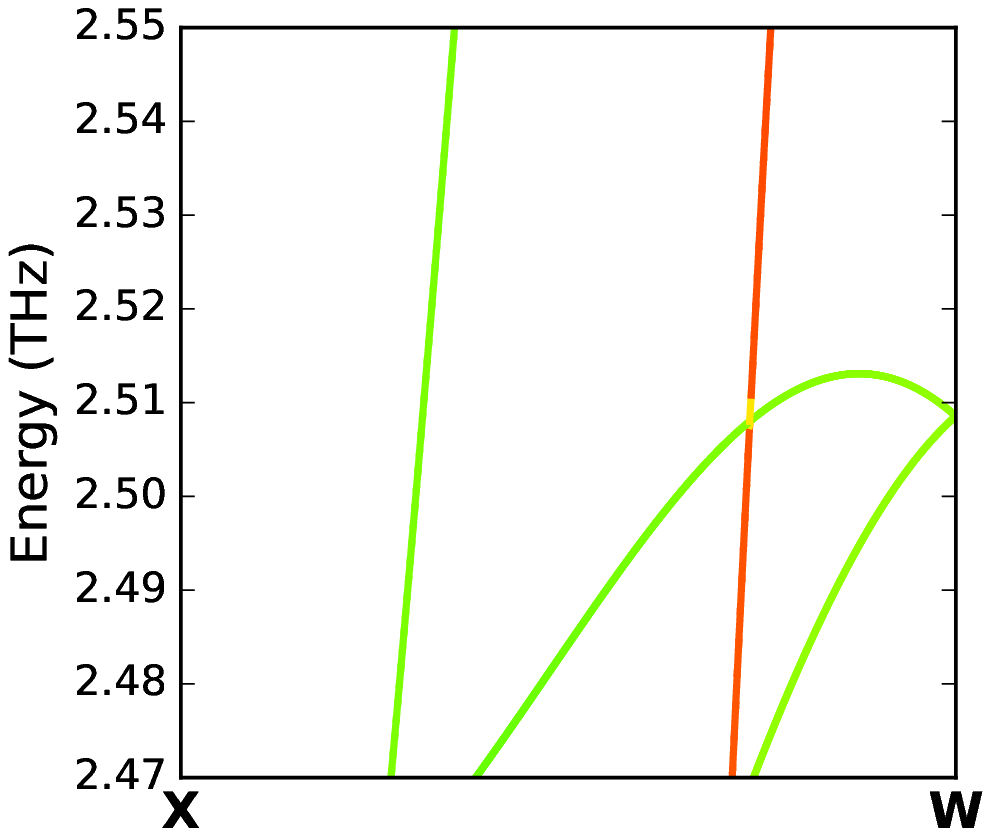}
   \caption{}
   \label{fig:}
\end{subfigure}
\caption{(Colour online) (a) Phonon dispersion plot of SnS along high-symmetric direction, phonon dispersion plot along with atomic contribution for box 1 of (b) SnS and box 2 of (c) SnS, (d) SnSe and (e) SnTe along X-W direction.}
\end{figure}

\begin{figure}[]
\begin{subfigure}{0.85\linewidth}
   \includegraphics[width=0.95\linewidth, height=6.5cm]{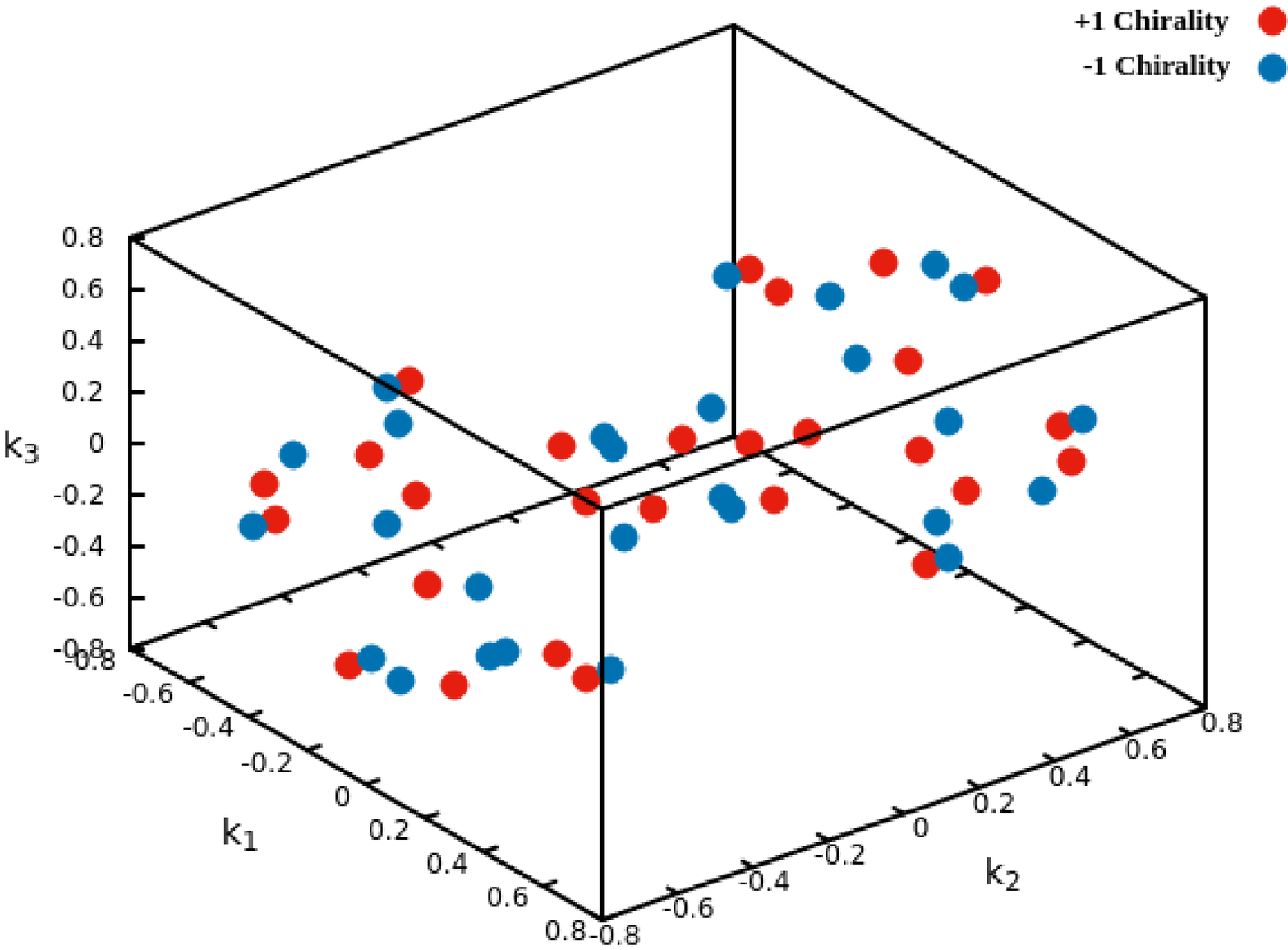}
   \caption{}
   \label{fig:}
\end{subfigure}
\begin{subfigure}{0.95\linewidth}
   \includegraphics[width=0.85\linewidth, height=5.8cm]{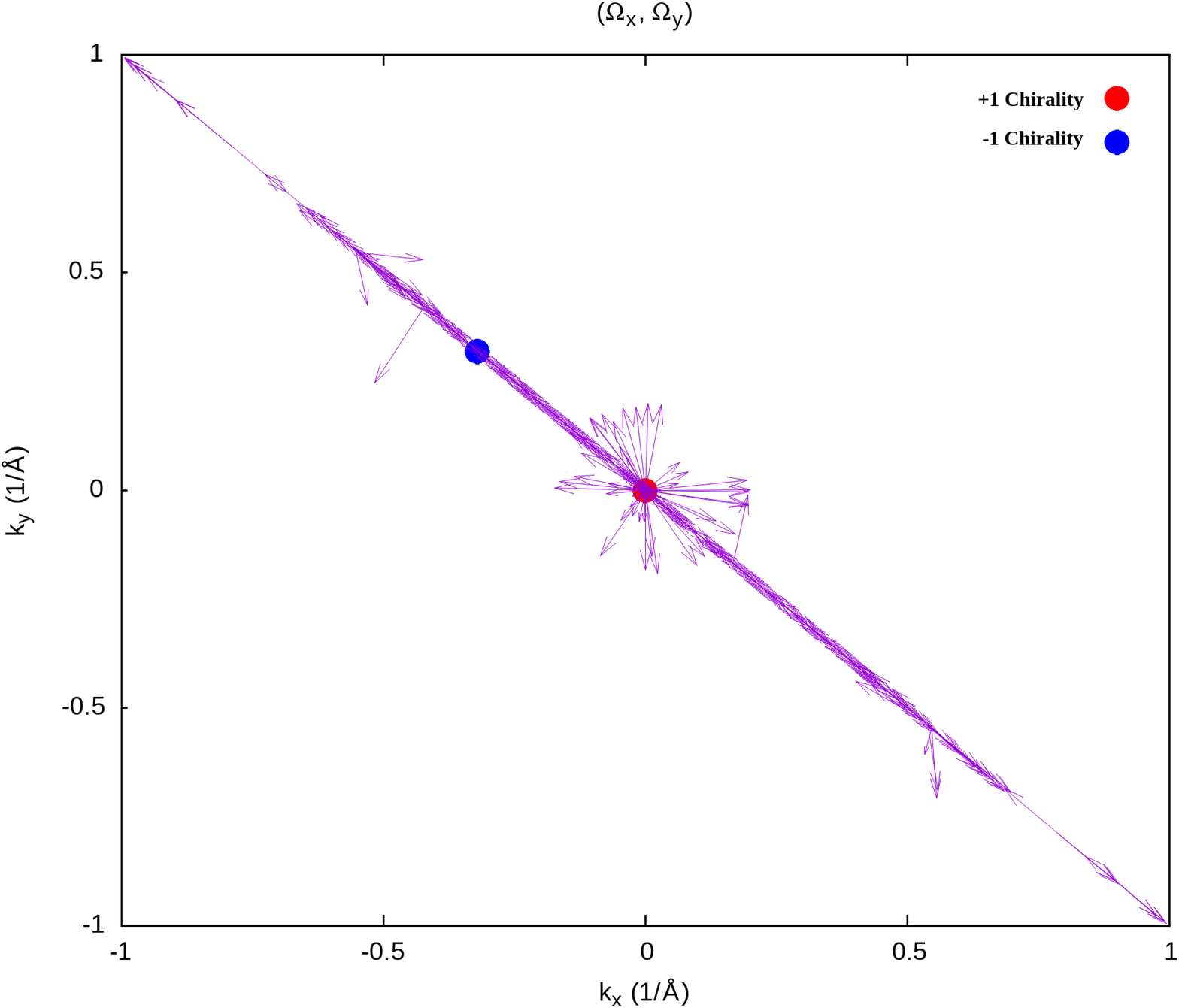}
   \caption{}
   \label{fig:}
\end{subfigure}
\caption{(Colour online) (a) The distribution of WPs for the first BZ due to the touching point 2 as shown in Fig. 1(a) and (b) Normalized Berry curvature of SnS.}
\end{figure}
  
  \textit{Computational details :-} WIEN2k code is used for carrying out the first-principle calculation using PBEsol exchange-correlation (XC) functional \cite{wien2k,pbesol}. The optimized lattice parameters of 5.753 \AA, 5.955 \AA \, and 6.291 \AA \, are utilized for SnS, SnSe and SnTe, respectively \cite{antik_jpcm,antik_jpcm2}. $F$m$\bar{3}$m space group, which denotes the rock-salt phase of these compounds, have been chosen for the present calculation along with the Wyckoff positions at (0.0, 0.0 ,0.0) and (0.5, 0.5, 0.5) for Sn and X (= S, Se, Te), respectively. The 2$\times$2$\times$2 supercell is taken for performing the force calculation by setting the force convergence criteria at 0.1 mRy/Bohr. The non-analytical term correction (NAC) is added for the entire phonon calculation to lift the degeneracy of longitudinal optical (LO) and transverse optical (TO) band at $\textbf{q}$ $\rightarrow$ 0 \cite{togo_nac}. PHONOPY code is used for the phonon calculation, where the real-space interatomic force constant ($\Phi_{\alpha \beta } (\mathbf{R}_{jl} - \mathbf{R}_{j'l'})$) is estimated through finite-displacement method \cite{phonopy}. In order to investigate the non-trivial topological properties of phonon modes, the tight-binding Wannier Hamiltonian for these vibrational modes are formed using $\Phi_{\alpha \beta } (\mathbf{R}_{jl} - \mathbf{R}_{j'l'})$, which are given by \cite{liu_2017}, 
  
\begin{eqnarray}
H = \frac{1}{2} \sideset{}{}\sum_{l,m,\alpha} \sideset{}{}\sum_{l',m',\beta} \Phi_{\alpha \beta } (\mathbf{R}_{jl} - \mathbf{R}_{j'l'}) u^\alpha_{lm}  u^\beta_{l'm'}
\end{eqnarray}

where, $\alpha , \beta $ represent Cartesian indices, $u^\alpha_{lm}$ ($u^\beta_{l'm'}$) denotes the displacement component of the $l$th ($l'$th) atom in the $j^ {th}$ ($j'^{ th}$) unit cell along the direction of $\alpha$ ($\beta$). All the topological properties of the phonons are explored using WANNIERTOOLS code \cite{wanniertools}.

\textit{Results and discussions :-} Phonon dispersion curve of SnS along the high-symmetric direction is shown in Fig. 1(a), whereas the phonon band structures of SnSe and SnTe are plotted in Figs. S1(a) and S1(b) of Supplemental Material \cite{supple}. For present study, NAC is included to compute all phonon's related calculations along with the topological properties. The detailed description of the phonon band structures for all these materials are already discussed in different literatures \cite{antik_pla,asihi_sse}. Three optical and three acoustic phonon branches are obtained for these compounds. Some of the optical branches have shown linear touching along X-W direction as marked by red boxes and have denoted by numbers 1 and 2, respectively, in Fig. 1(a). Now, in order to observe the presence of phononic band inversion as compared to electronic topological materials, these two touching points of SnS are plotted separately for seeing the clear view of band touchings in Figs. 1(b) and (c), where red (green) colour denotes the contribution of Sn (X = S, Se and Te) atom to form this band around the observed energy window. Fig. 1(b) shows the absence of band inversion at touching point 1. But, in case of point 2, the band inversion between phonon branches of Sn and S atoms is clearly seen from Fig. 1(c). In such case, the phonon branches are characterized by the corresponding atoms using projected density of states of individual atom. The linear touching of two optical branches at the point 2 along X-W directions shows slightly tilted nature, which provides a convincing reason to classify this point as type-II Weyl phononic point at $\sim$2.83 THz. For comparison purpose, similar touching point between the corresponding two bands at point 2 for SnSe and SnTe are shown in Figs. 1(d) and (e), respectively. These touching points for SnSe and SnTe are found to be $\sim$2.46 THz and $\sim$2.51 THz, respectively. The presence of phononic band inversion is evident from these figures for this particular point for all the samples. Moreover, the nature of tilted band touching is more profoundly seen when one moves from SnS to SnTe. This behaviour is nicely matched with the previously predicted ideal type-II Weyl phononic material ZnSe \cite{liu_103}.     

\begin{figure}[]
\begin{subfigure}{0.95\linewidth}
   \includegraphics[width=0.85\linewidth, height=4.8cm]{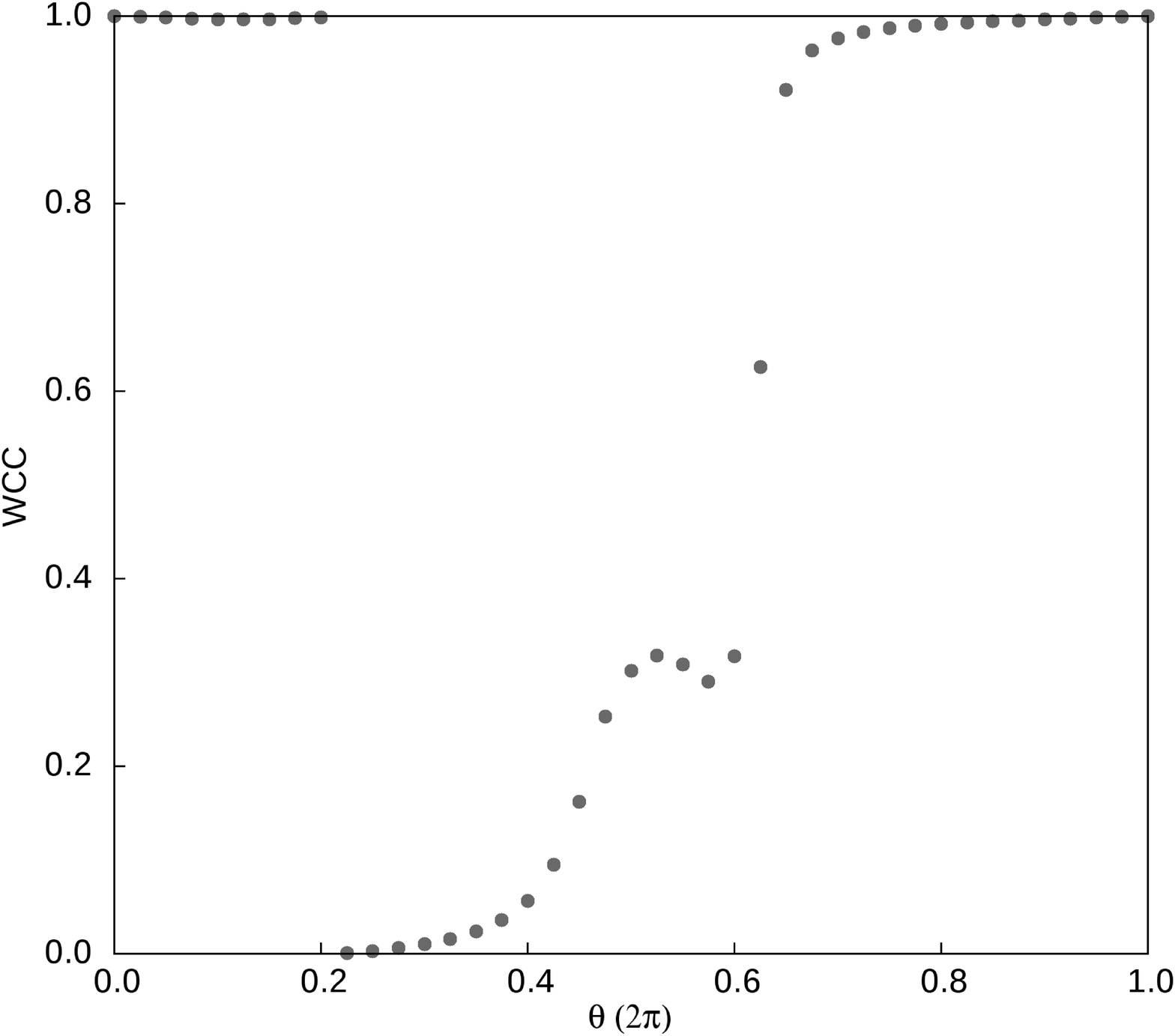}
   \caption{}
   \label{fig:}
\end{subfigure}
\begin{subfigure}{0.95\linewidth}
   \includegraphics[width=0.85\linewidth, height=4.8cm]{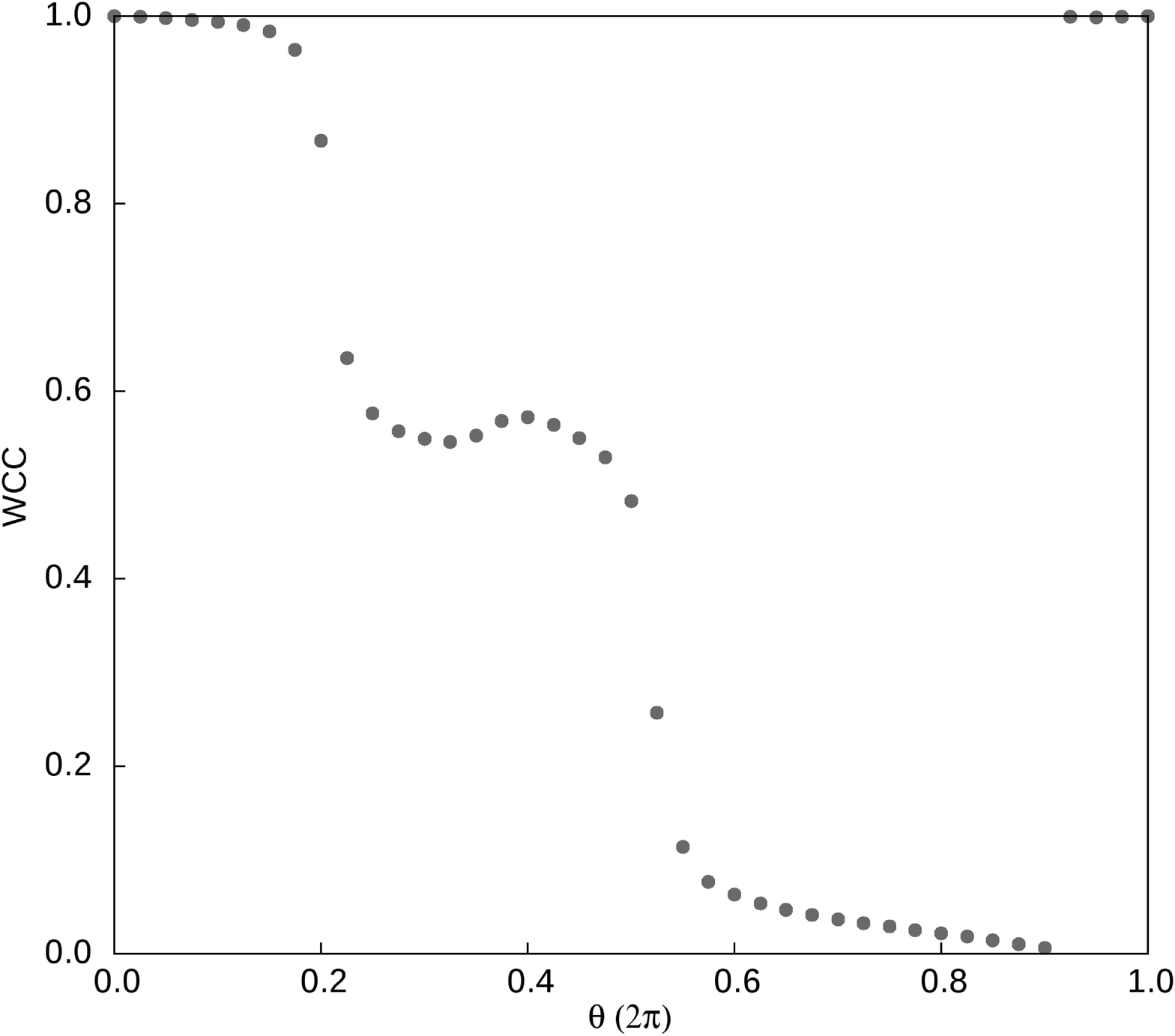}
   \caption{}
   \label{fig:}
\end{subfigure}
\caption{(Colour online) Wannier charge center evolution of SnS around the (a) positive and (b) negative charges of a pair of Weyl points.}
\end{figure}

  We have found 56 WPs in full Brillouin zone (BZ) for SnS after applying all symmetry operations, where these points are shown in Fig. 2(a). Here, 28 points have shown $\mathscr{C}$ = +1 and the remaining 28 points have been found to estimate the value of $\mathscr{C}$ = -1. In case of SnSe and SnTe, we have estimated the values of 24 and 54 numbers of WPs, respectively, where the $\mathscr{C}$ of the corresponding sample is found to be conserved. In order to explain the topological properties of these WPs using the phonon bands, the Berry curvature needs to be calculated using the following relation \cite{wanniertools}, 

\begin{eqnarray}
\Omega_n^z (\mathbf{q}) = \nabla_\mathbf{q} \times i\langle u_n (\mathbf{q})|\nabla_\mathbf{q}|u_n (\mathbf{q})\rangle 
\end{eqnarray}  
  
where $u_n (\mathbf{q})$ denotes the phonon branches of $n^{th}$ band. It is known that one pair of Weyl nodes typically contains the positive and negative Berry curvatures. In momentum space, these positive and negative Berry curvatures give the information of source and sink of the Berry flux, respectively. For SnS, Fig. 2(b) illustrates the Berry curvature of crossing nodes in $\mathbf{q}_z$ = 0 plane. In the figure, the source and sink of the Berry flux are mentioned by the red and blue points, respectively. Further, to get the evident of positive and negative chiral charges around the WPs, the Wannier charge centers (WCC) are calculated and plotted in Figs. 3(a) and (b), respectively, for SnS. Here, WCC is calculated using the Wilson loop approaches around the corresponding Weyl point. Now, it is noted that the Fig. 3(a) (3(b)) shows the Wannier charge evolution around one particular Weyl node, which shows positive (negative) chirality. This behaviour is found to be similar with the previously predicted type-II Weyl phononic materials \cite{xia_cdte,liu_103}. SnSe and SnTe have also shown the similar kind of behaviour as compared to SnS. However, the comparative study of phononic topological properties for these compounds needs to be explored further.

\begin{figure}
  \begin{center}
    \includegraphics[width=0.95\linewidth, height=8.0cm]{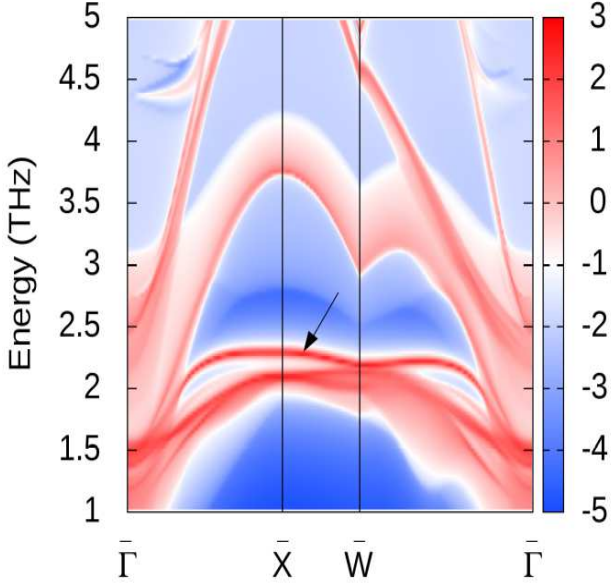} 
    \caption{(Colour online) The surface local density of states for the (001) surface of SnS along two dimensional high-symmetric directions. The surface arc is marked by a black arrow.}
   \label{fig:}
   \end{center}
\end{figure} 

\begin{figure}[]
    \begin{subfigure}{0.45\linewidth}
   \includegraphics[width=0.95\linewidth, height=3.5cm]{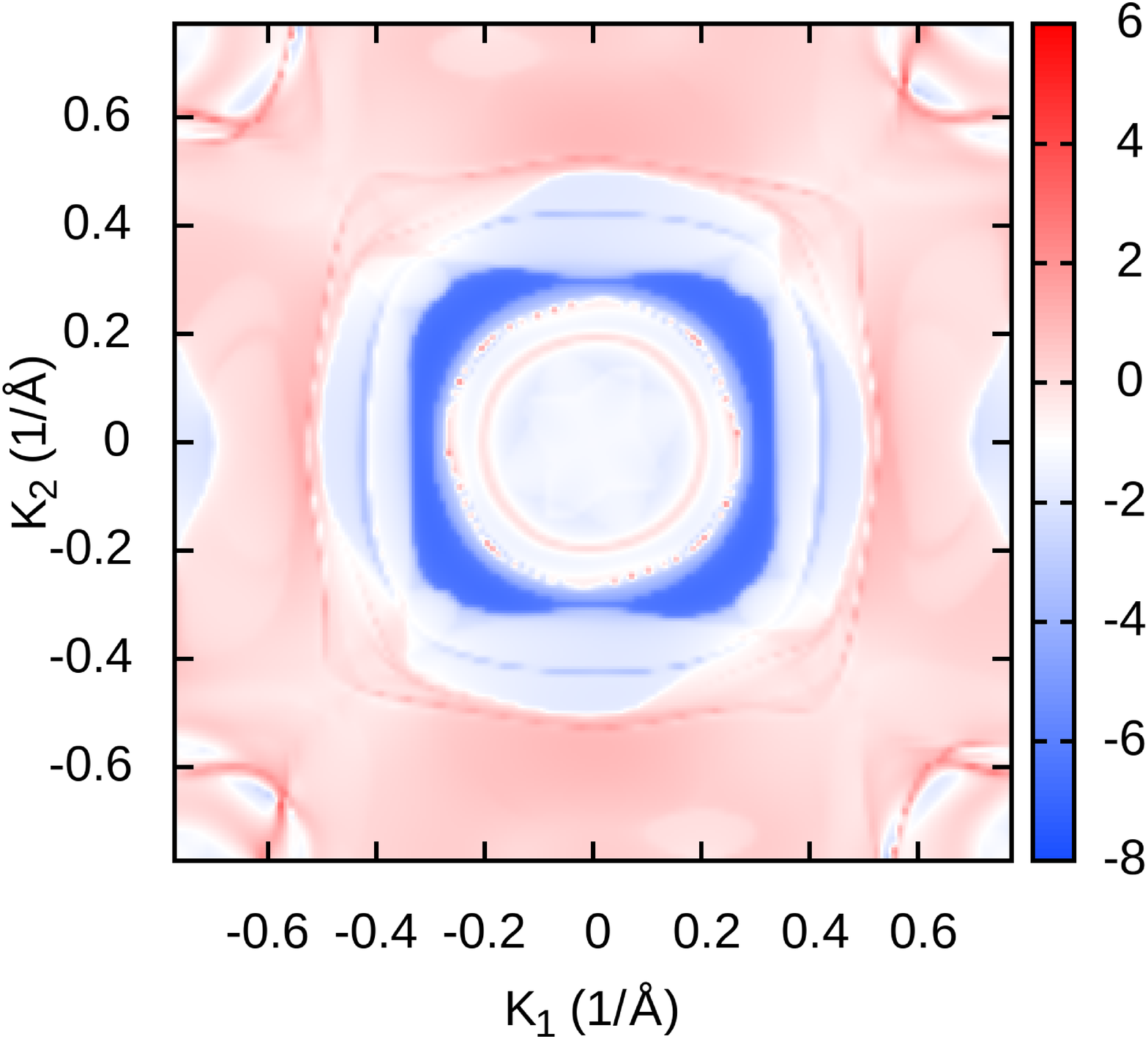}
   \caption{Energy = 2.0 THz}
   \label{fig:}
\end{subfigure}
\begin{subfigure}{0.45\linewidth}
   \includegraphics[width=0.95\linewidth, height=3.5cm]{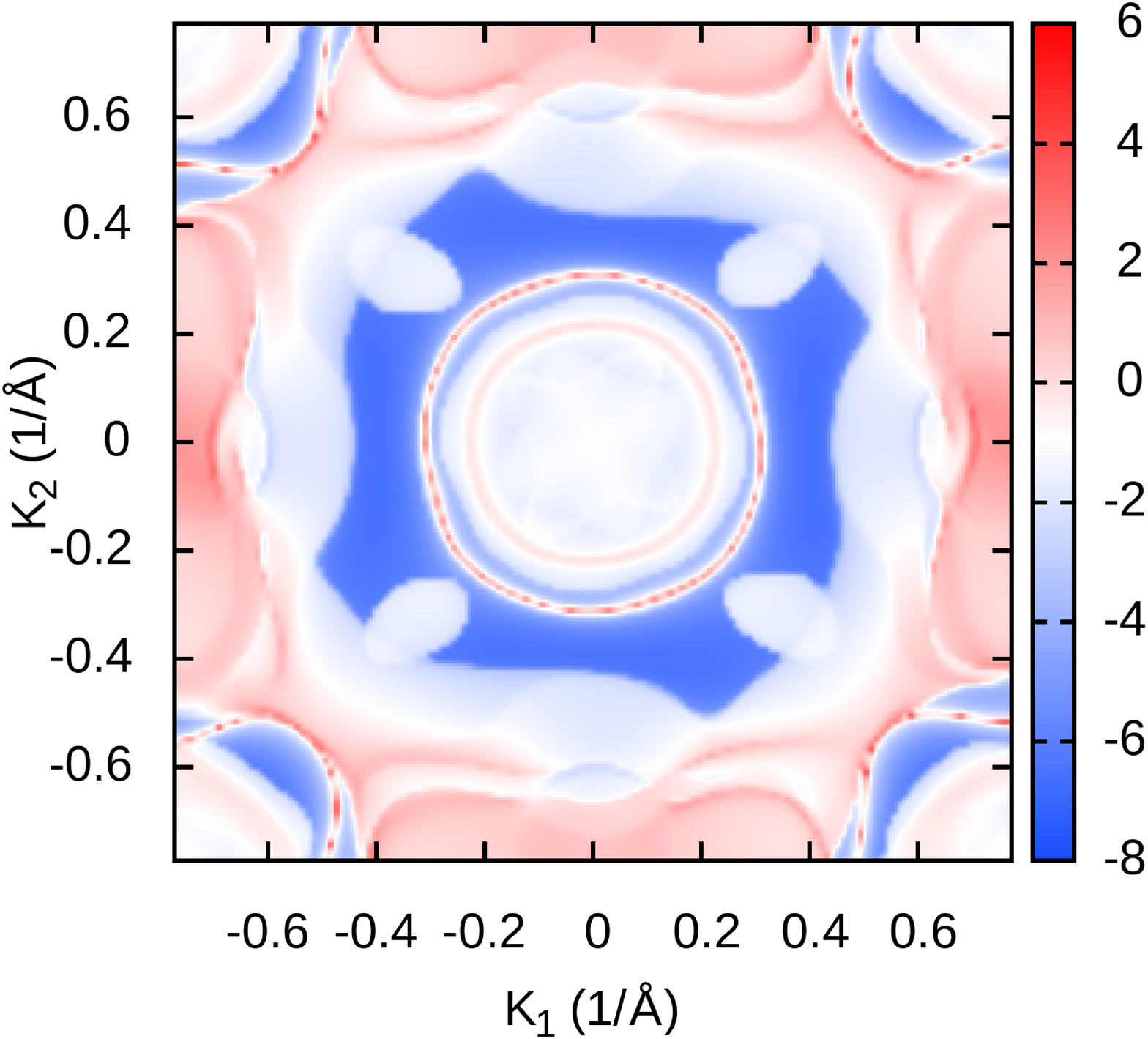}
   \caption{Energy = 2.1 THz}
   \label{fig:}
\end{subfigure}
\begin{subfigure}{0.45\linewidth}
   \includegraphics[width=0.95\linewidth, height=3.5cm]{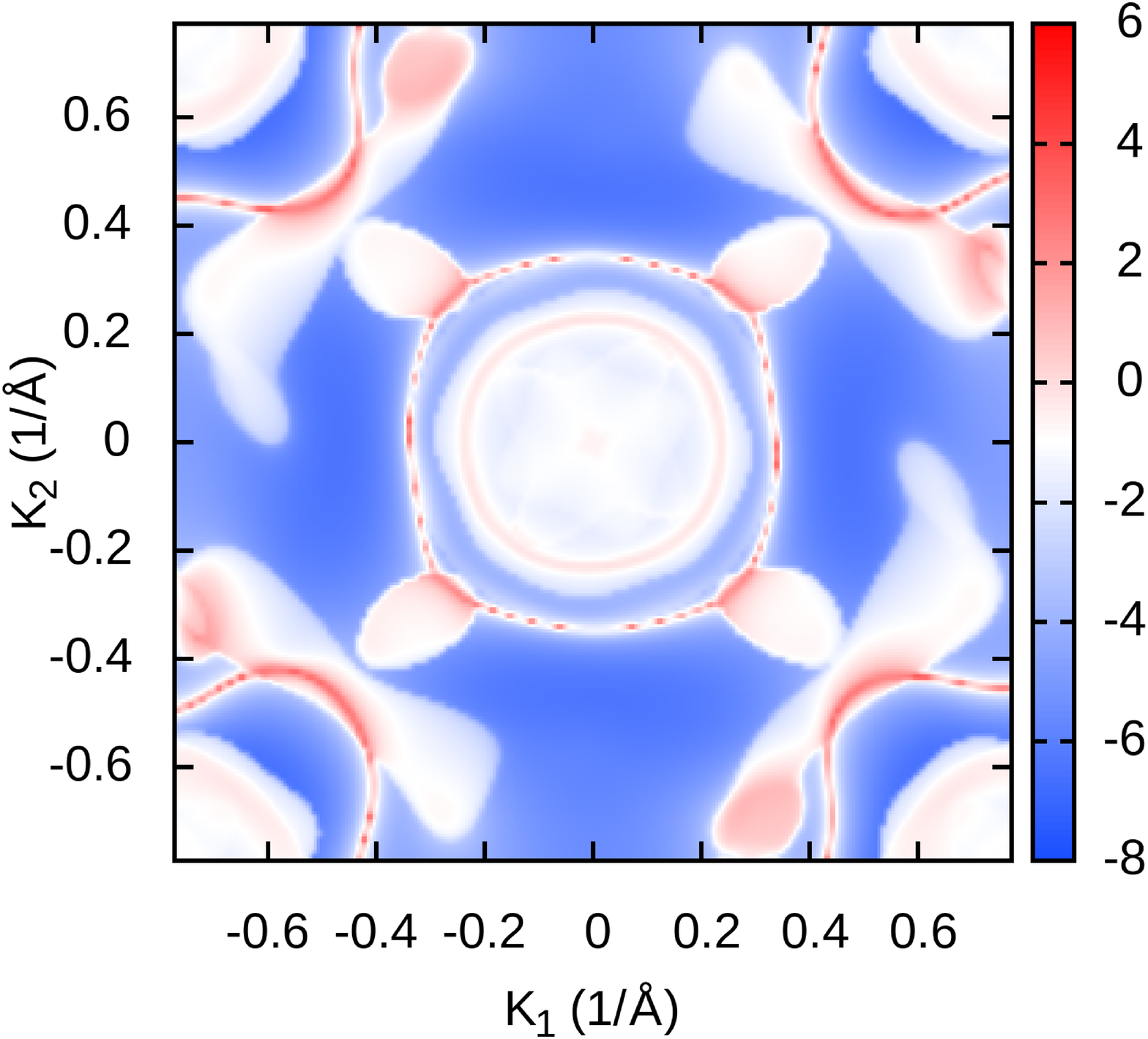}
   \caption{Energy = 2.15 THz}
   \label{fig:}
\end{subfigure}
\begin{subfigure}{0.45\linewidth}
   \includegraphics[width=0.95\linewidth, height=3.5cm]{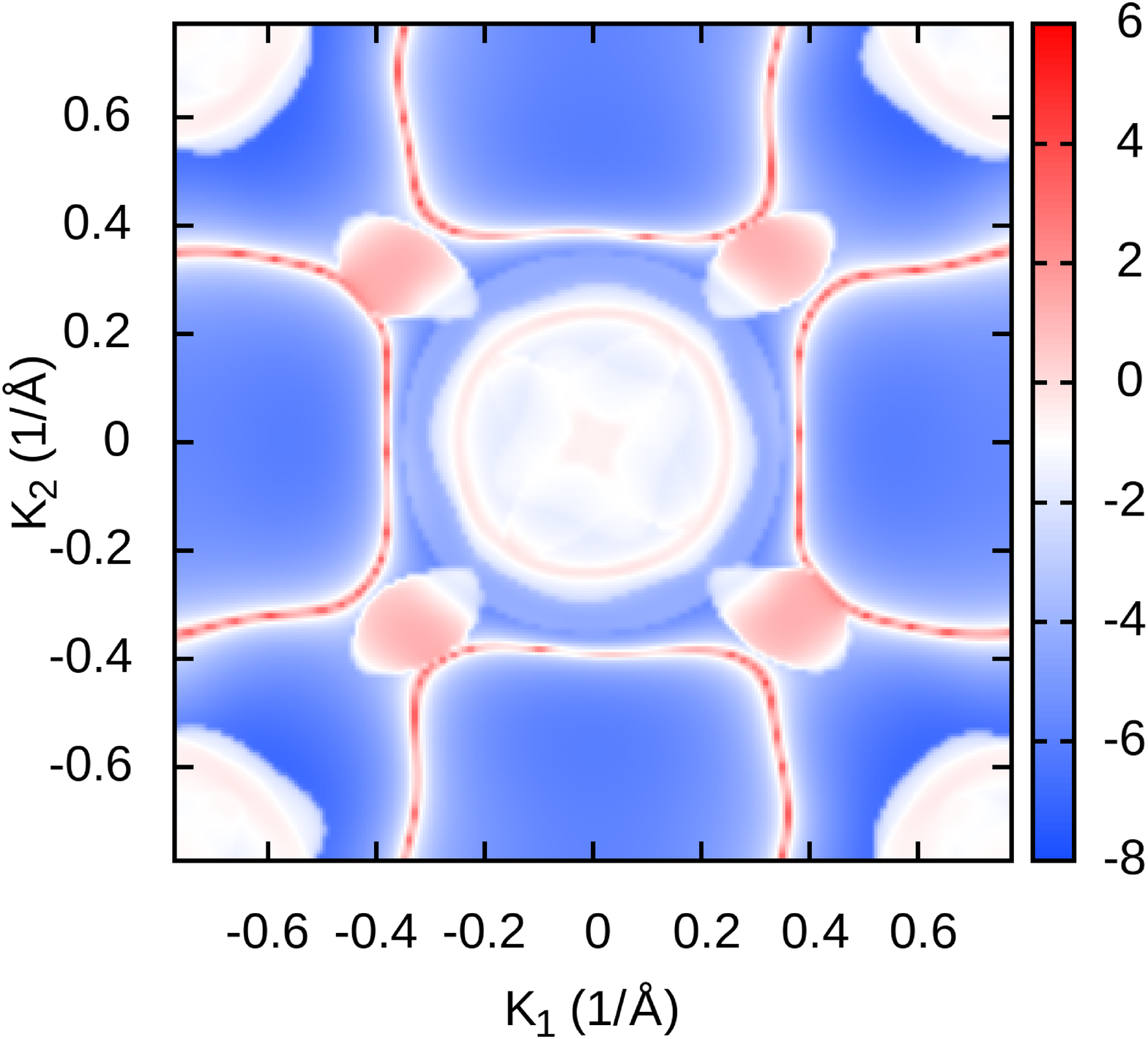}
   \caption{Energy = 2.2 THz}
   \label{fig:}
\end{subfigure}
\begin{subfigure}{0.45\linewidth}
   \includegraphics[width=0.95\linewidth, height=3.5cm]{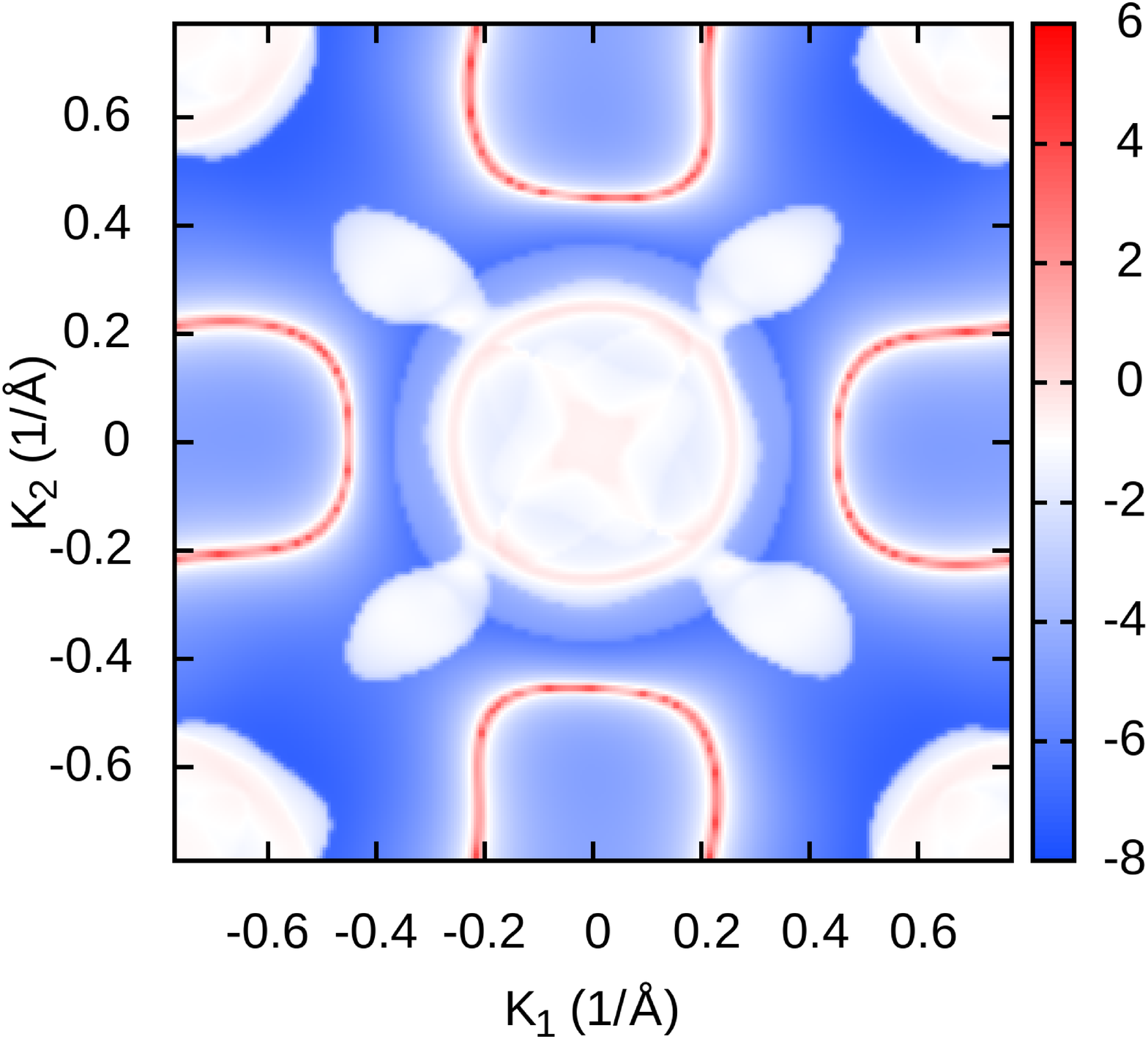}
   \caption{Energy = 2.25 THz}
   \label{fig:}
\end{subfigure}
\begin{subfigure}{0.45\linewidth}
   \includegraphics[width=0.95\linewidth, height=3.5cm]{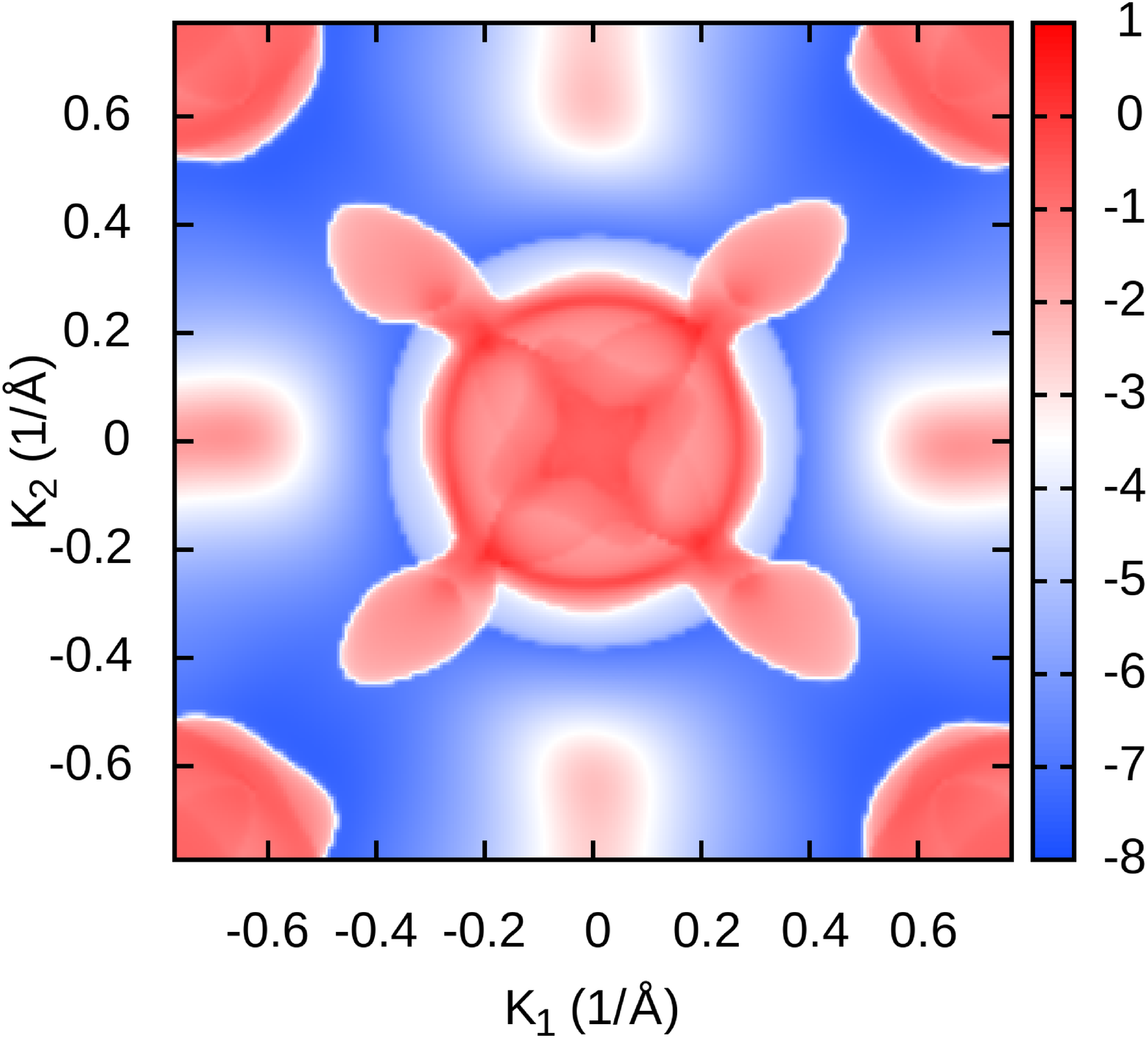}
   \caption{Energy = 2.3 THz}
   \label{fig:}
\end{subfigure}
\caption{(Colour online) The topologically protected non-trivial topological phononic surface states of SnS of (001) surface at different frequencies for the top surface of the unit cell.}
\end{figure}

  Now, in order to investigate the exotic topological phononic surface state for (001) surface, the surface local density of states (LDOS) of SnS is calculated along the surface high-symmetric direction, which is shown in Fig. 4. Moreover, the surface LDOS of SnSe and SnTe are shown in Supplemental Material \cite{supple}. To calculate the surface LDOS, the second rank tensor of force constant is estimated from the corresponding phonon calculation. Using this, we have obtained the tight-binding parameters for the bulk and surface atoms to study the topological behaviours of phonon. Afterwards, the surface spectrum is calculated by computing the surface Green's function (SGF) iteratively with considering the imaginary part of the SGF \cite{surface_green}. The surface BZ of (001) of $F$m$\bar{3}$m spacegroup is a square. It is known that the (001) surface shows mirror symmetry with respect to (1$\bar{1}$0) mirror plane \cite{tanaka_snte,y_sun}. The high-symmetric points of 3D momentum space are projected on (001) surface. Therefore, the $\Gamma$, X and W points are now denoted by $\bar{\Gamma}$, $\bar{X}$ and $\bar{W}$, respectively. We have chosen this direction because the linear touching is observed along X-W direction in 3D momentum space. It is clearly seen from the figure that the presence of surface arc is evident along $\bar{\Gamma}$ - $\bar{X}$ - $\bar{W}$ - $\bar{\Gamma}$ direction around 2.0 - 2.2 THz energy region, which is also marked by an black arrow. The similar behaviour is found for FeSi, which is a famous candidate for a topological phononic material \cite{tzhang}. For SnSe and SnTe, we have seen the similar kind of surface arc, which are shown in Fig. S2(a) and S2(b) of Supplemental Material, respectively \cite{supple}. This behaviour of all the materials ($i.e.$ SnX (X = S, Se, Te)) is indicating the presence of type-II Weyl topological phonon. Moreover, the isofrequency surface contours provides the information of Fermi arc at particular frequency. Therefore, we have also computed isofrequency surface states, which have plotted in Figs. 5(a), (b), (c), (d), (e) and (f) at energy 2.0 THz, 2.1 THz, 2.15 THz, 2.2 THz, 2.25 THz and 2.3 THz, respectively. Here, the isofrequency surface states are shown for SnS material for the top surface of the unit cell, whereas the projection of bulk on $k_1$-$k_2$ plane and the bottom surface of the unit cell are shown in Figs. S3 and S4 of Supplemental Material  \cite{supple} at the different aforementioned frequencies. At this point, it is important to note that if the bulk projection and any surface ($i.e.$ top/bottom) projection are showing similar behaviour then the Fermi arc due to topological behaviour of the material is not present at this particular energy. In present scenario, the evidence of Fermi arc is found at energy 2.2 THz and 2.25 THz because the surface projection and the bulk projection are showing different spectrum. The similar behaviour is also seen for SnSe and SnTe materials. Therefore, these three samples ($i.e.$ SnS, SnSe and SnTe) in rock-salt phase are predicted as the candidates of type-II Weyl topological phononic materials. The present theoretical study can be possible to verify by performing the corresponding experiments on these materials. Moreover, this predicted result enhances the usefulness of SnS, SnSe and SnTe materials for practical application purposes.         
   
\textit{Conclusions :-} In this present study, the topological behaviours of phonon are explored for SnS, SnSe and SnTe materials in rock-salt phase. The tilted linear band touching is seen from phonon dispersion curves of all the materials along X-W direction. The band inversion around this point are also investigated for all the materials, where the frequency value of this topological point for SnS is found to be $\sim$2.83 THz. In case of SnSe and SnTe, this topological points are seen at $\sim$2.46 THz and $\sim$2.51 THz, respectively. The computed values of WPs are 56, 24 and 54 for SnS, SnSe and SnTe materials, respectively, in the full BZ. The values of Chiral charges, which are present at these WPs of corresponding materials, are estimated to be conserved. In case of all the materials, the surface arc is clearly seen from the surface local density of states. Also, the presence of Fermi arc is evident from the isofrequency surface state of individual compounds. For SnS, the Fermi arcs are found at $\sim$2.2THz and $\sim$2.25 THz frequencies. All these behaviours are suggested to predict SnS, SnSe and SnTe as type-II Weyl phononic materials. The present study reveals the topological importance in phononic structure of these famous electronic topological materials, which will enrich the practical applications of all these materials.

\end{document}